\documentclass[twocolumn,superscriptaddress,floatfix]{revtex4-1}

\usepackage{graphicx}%
\usepackage{dcolumn}%
\usepackage{bm}%
\usepackage{multirow}

\usepackage{color}
\usepackage{comment}

\newcommand{\rev}[1]{\textcolor{black}{#1}}

\begin{document}

\title{
First-principles study of exchange coupling constants in Nd$_x$Fe$_{1-x}$ amorphous alloy
}
\author{Asako Terasawa}
\email{a.terasawa@rist.or.jp}
\thanks{
Present affiliation: Research Organization for Information Science and Technology,
Japan.
}
\author{Yoshihiro Gohda}
\email{gohda.y.ab@m.titech.ac.jp}
\affiliation{
Department of Materials Science and Engineering, Tokyo Institute of Technology,
J1-3, Nagatsuta-cho 4259, Midori-ku, Yokohama 226-8502, Japan}
\date{}%

\begin{abstract}
We investigate the exchange coupling constant $J_{ij}$ in Nd$_x$Fe$_{1-x}$ amorphous alloys with different compositions within the framework of first-principles calculation.
We observed a strong atomic-dependence of $J_{ij}$ and its fluctuations. 
We show that the composition strongly affects the distance dependence of $J_{ij}$.
Composition dependence of calculated Curie temperatures is modest for $x<0.5$.
To examine the effect of the local environment on the exchange couplings, we demonstrate combined analyses of the coordination structure and exchange coupling constants using the Gabriel graph.
Our study reveals that the Curie temperatures are mostly dominated by the averaged $J_{ij}$ and coordination numbers determined by the pairs of neighboring atoms.
We also observed that the exchange couplings between Fe--Fe and Fe--Nd become stronger with increasing the number of surrounding Nd atoms.
\end{abstract}

\maketitle

\section{Introduction}\label{sec:intro}

Development of a high-performance Nd--Fe--B permanent magnets is one of the most challenging problems in achieving the high-energy efficiency of next-generation vehicles and turbines. The role of the non-magnetic grain boundary (GB) phase was intensively studied in recent years \cite{S_Sugimoto_2011, K_Hono_2012, S_Hirosawa_2017, S_Li_2002, W_F_Li_2009, T_H_Kim_2012, U_M_R_Seelam_2016}, 
because the magnetic insulation between main phase grains dominates the prevention of magnetization reversal at the GB phases \cite{K_Hono_2012,H_Sepehri_Amin_2012}.
Moreover, a curious feature of the GB phase between two main phase grains was reported, where both the crystallinity and composition ratio of the GB phase vary depending on the relative angles of neighboring main phase grains with the $c$ axis \cite{T_T_Sasaki_2016,X_D_Xu_2018}. 
They also found that the amorphous GB phases appeared at the interfaces parallel to the $c$ axis whereas the crystalline GB phases appeared perpendicular to the $c$ axis of the Nd$_2$Fe$_{14}$B grains with the Nd composition ratio of approximately 40\%, and 60\%, respectively.

Measuring the magnetic properties of actual grain boundary phases is very difficult because of their structural complexities and their sizes of approximately 1--2 nm. First principles calculation technique appears to be a promising tool to determine the magnetic properties of the complicated microstructures because it can determine the electronic structures of atomistic models without the need to perform experiments.
Until now, several efforts have been made to calculate the magnetic properties of various phases and their interfaces in permanent magnets \cite{ A_Saengdeejing_2016, Y_Tatetsu_2016, Z_Torbatian_2016, N_Umetsu_2016, Y_Gohda_2018, Y_Tatetsu_2018,  A_M_Schonhobel_2019}.
However, it is little known about the actual exchange couplings at the GB phases despite the extensive examination of the exchange coupling constants of main phases done so far using first principles calculation techniques \cite{T_Fukazawa_2017, H_Akai_2018, C_E_Patrick_2018, T_Fukazawa_2019, C_E_Patrick_2019}. 
This is partly because the structures of the GB phases are very complicated and not well understood, as described in Ref.~\cite{T_T_Sasaki_2016}.
Very recently, an attempt has been made to calculate the exchange coupling constants in the crystalline Nd--Fe alloy as a candidate of crystalline GB phase of the Nd--Fe--B permanent magnet \cite{Y_Ainai_2020,Y_Ainai_2020_2}.
Results from this study strongly suggest that the crystalline Nd--Fe alloy is ferromagnetic, even though its Nd composition ratio is up to 67\%.

In this study, we examined the exchange coupling constants of amorphous Nd$_{x}$Fe$_{1-x}$ alloys for a wide range of compositions in the framework of spin-dependent density functional theory.
We obtained strongly fluctuating $J_{ij}$ curves which changes the shape depending on the value of $x$. By comparing the $J_{ij}$ curves for different compositions, we observed that the $J_{ij}$ curves became steeper for large $x$.
Moreover, we performed a combined analysis to examine the relationship between the exchange coupling constants with the local structures, using Gabriel graph analysis for amorphous systems \cite{A_Terasawa_2018}.
We found that the average $J_{ij}$ for neighboring atom pairs became larger with an increase in the $x$ value, 
which resulted in the modest decrease of the calculated Curie temperature of amorphous Nd$_{x}$Fe$_{1-x}$ depending on $x$.
We also found that the exchange coupling constants vary depending on the local environment, and the presence of Nd atoms in the circumstances enhances the exchange coupling constants between Fe--Fe and Fe--Nd.

\begin{figure*}[t]
\begin{center}
\includegraphics[width=17 cm]{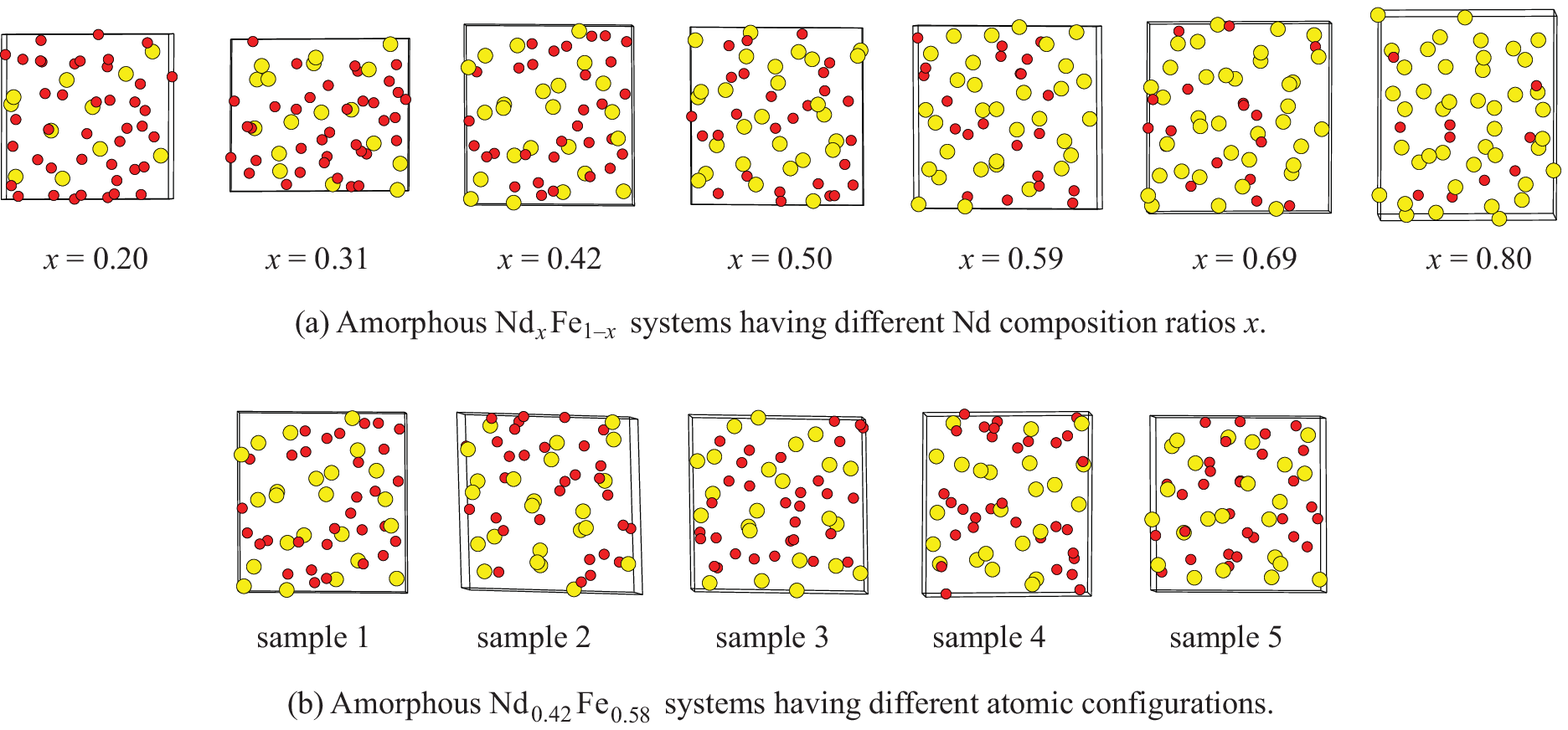}
\end{center}
\caption{Examples of amorphous Nd$_{x}$Fe$_{1-x}$ systems created by melt-quench simulations. In the figures, the large yellow circles, and the small red circles correspond to the Nd and Fe atoms, respectively. In the panel (a), systems with different compositions are compared. In panel (b), systems with $x=0.42$ having different atomic configurations are compared.
}
\label{fig:samples}
\end{figure*}

\section{Computational models and methods}

We performed the computations in this paper within the framework of density functional theory.
For the density functional calculations, we used OpenMX code \cite{T_Ozaki_2003}. We adopted the Perdew-Burke-Ernzerhof exchange-correlation functional \cite{GGA-PBE} with the generalized gradient approximation (GGA-PBE) unless otherwise stated. 
Cutoff radii were set to 6.0 Bohr for Fe and 8.0 Bohr for Nd. We adopted the pseudopotentials by the Morrison-Bylander-Kleinman scheme \cite{MBK},
which takes into account the spin-orbit coupling and the $4f$ state of Nd as a spin-polarized core state by considering the occupation of three electrons.
The convergence criteria for the force and the total energy were chosen as $1.0\times10^{-3}$ Hartree/Bohr and $1.0\times10^{-6}$ Hartree, respectively.

To determine the atomic structures of Nd$_{x}$Fe$_{1-x}$ alloys, we adopted the melt quench method based on the spin-independent first-principles molecular dynamics (FPMD) simulations.
To prepare the atomic coordinates of Nd--Fe amorphous alloys, we adopted the melt-quench molecular dynamics method \cite{MQ1} as follows.
A crystalline alloy having 27 atoms of each Fe and Nd was prepared in the first step.
Then, a certain number of atoms were substituted to get the desired compositions for 7 different systems with $x=0.20,~0.31,~0.42,~0.50,~0.59,~0.69$ and $0.80$.
The alloys were melted at 4000 K for 1 ps, quenched to 300 K in 2 ps, and then stabilized at 300 K for 2 ps by the FPMD simulations.
A structural optimization regarding the internal coordinates and the lattice vectors was applied after the melt-quench procedure for each composition.
Next, the systems were annealed at 900 K for 1 ps, cooled to 300 K in 2 ps, and then stabilized at 300 K for 2 ps, which was followed by the second structural optimization.
Finally, the procedures of melting-quenching-stabilizing, annealing-cooling-stabilizing, and structural optimization were repeated to obtain 5 independent samples for each composition. 
In the procedures explained here, first-principles molecular dynamics simulations were performed with moderate accuracy because of the computational costs of FPMD. We used $1\times 1\times 1$ $k$-grid with the cutoff energy of 300 Ry and electronic temperature of 2000 K as computational conditions. For the pseudoatomic orbital basis sets, we adopted s1p1d1 and the s2p1d1 basis set for Fe and Nd, respectively, where $3p$ states of Fe and $5s$ and $5p$ states of Nd were treated explicitly as valence states.
The time step $\Delta t$ was set as 1 fs. For the finite temperature FPMD simulations, we adopted the velocity scaling method.

After the above procedures, we again performed the structure optimization with higher computational accuracy in order to obtain more stable structures of amorphous alloys with considering the magnetic interactions.
In this process, spin-dependent density functional calculations were performed with $7\times 7\times 7$ $k$-grid with a cutoff energy of 500 Ry and electronic temperature of 300 K. For the pseudoatomic orbital basis sets, we adopted s2p2d2 and the s3p2d2 basis sets for Fe and Nd, respectively, which 
implied more accurate calculation conditions than that in the FPMD simulations. 
In the spin-dependent density functional calculations, we assume 
the spin population orientation to be upspin and downspin for Fe and Nd, respectively.
Figure \ref{fig:samples} shows some of the resultant amorphous Nd$_{x}$Fe$_{1-x}$ systems in this study. It is possible to see in Fig.~\ref{fig:samples}(b) that different samples with the same composition have non-identical atomic configurations.

For the computation of exchange coupling constants of the simulated systems, 
we adopted the Liechtenstein method \cite{A_I_Liechtenstein_1987} combined with the first-principles calculation results by OpenMX.
To implement the Liechtenstein method applicable to OpenMX, we adopted the finite pole approximation of Fermi distribution function \cite{T_Ozaki_2007, A_Terasawa_2019}, and developed the single-site orthogonalization scheme, which allows calculating exchange coupling constants of rare earth metals with correspondence to the experimental results \cite{A_Terasawa_2020}.

\section{Computational results}

\subsection{Exchange coupling constants of amorphous Nd$_{x}$Fe$_{1-x}$ alloys}

\begin{figure}[h]
\begin{center}
\includegraphics[width=6.5 cm]{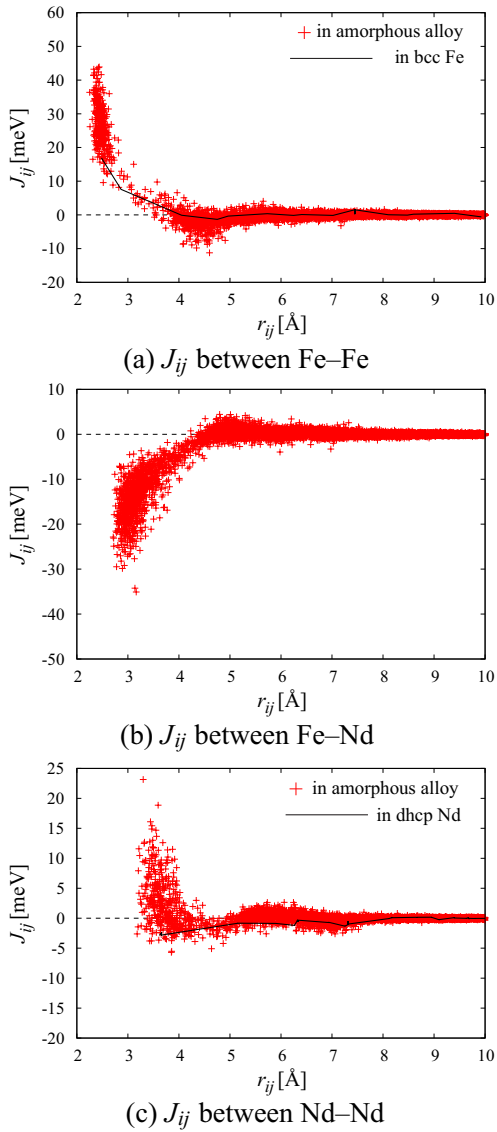}
\end{center}
\caption{Exchange coupling constants of amorphous Nd$_{0.42}$Fe$_{0.58}$ alloys. In each panel, the $J_{ij}$ is plotted against $r_{ij}$ is represented by the red crosses. In panel (a) and (c), the $J_{ij}$s of body-centered cubic (bcc) Fe and double hexagonal close-packed (dhcp) Nd crystals are shown using the black solid lines.
}
\label{fig:Jij_42}
\end{figure}

Figure \ref{fig:Jij_42} shows calculated exchange coupling constants $J_{ij}$ for $x = 0.42$ amorphous alloys as functions of atomic distances between (a) Fe--Fe, (b) Fe--Nd, and (c) Nd--Nd. It is possible to see in each panel of Fig.~\ref{fig:Jij_42} that the $J_{ij}$ decays with the atomic distance but fluctuates strongly. 
In Fig.~\ref{fig:Jij_42}(a), we also show the exchange coupling constants of body-centered cubic (bcc) crystal of
Fe using a black solid line together with the $J_{ij}$ between Fe--Fe in the amorphous alloys.
Comparing the exchange coupling constants of amorphous Nd$_{x}$Fe$_{1-x}$ with that of bcc Fe, it is possible to see that the significant enhancement of $J_{ij}$ occurs at the smaller atomic distance than 2.48 \AA, which is the atomic distance of nearest neighbor pairs of bcc Fe.
While the enhancement of $J_{ij}$ depending on the atomic distance is so large, it is also necessary to look at the large fluctuation of $J_{ij}$
of the order of 10 meV.

A similar feature can be seen in Fig.~\ref{fig:Jij_42}(c), where the variation in $J_{ij}$ between Nd--Nd is plotted for both dhcp Nd and amorphous Nd$_{0.42}$Fe$_{0.58}$.
It is also noteworthy that mostly positive $J_{ij}$s between Nd--Nd are found for small atomic distances. 
The most significant positive $J_{ij}$s are found for the smaller atomic distances than 3.63 \AA, which is the atomic distance of nearest neighbor pairs of  double hexagonal close-packed (dhcp) crystal of Nd, while many positive values are found even for larger atomic distances.
This feature is different from the magnetism of dhcp Nd, where the nearest neighbor pairs exhibit weak negative exchange couplings.
This indicates that the nature of electrons in Nd--Fe alloys is different from that in dhcp Nd.
Moreover, we point out that the nearest neighbor $J_{ij}$ between Fe--Nd are approximately 10--30 meV, which is comparable to the nearest neighbor $J_{ij}$ for bcc Fe.
This indicates that the exchange interaction in Nd--Fe amorphous alloy is larger than what is expected from the magnetic properties of their constituting elements.

\begin{figure*}[t]
\begin{center}
\includegraphics[width=17 cm]{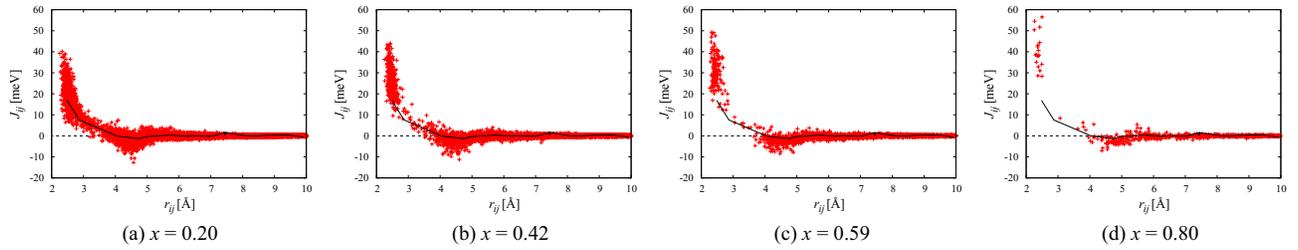}
\end{center}
\caption{Exchange coupling constants $J_{ij}$ between Fe--Fe as a function of atomic distance $r_{ij}$ for amorphous Nd$_{x}$Fe$_{1-x}$ alloys, for (a) $x=0.20$, (b) $x=0.42$, (c) $x=0.59$, and (d) $x=0.80$.
In each panel, the red crosses and the black solid line represent the $J_{ij}$ of amorphous alloy, and bcc Fe, respectively.
}
\label{fig:Jij_all}
\end{figure*}

The effects of the surrounding environment on $J_{ij}$ can be seen more evidently when it is compared among different compositions.
In Fig.~\ref{fig:Jij_all}, $J_{ij}$ between Fe--Fe are shown for (a) $x=0.20$, (b) $x=0.42$, (c) $x=59$, and (c) $x=80$.
We can see that the maximum values of $J_{ij}$ vary from 40 to 60 meV as $x$ increases from 0.20 to 0.80.
Moreover, there is a large difference between the $J_{ij}$ curves of bcc Fe and amorphous alloy for $x=0.20$ and $x=0.42$, while the difference is more subtle in the case of $x=0.59$ and $x=0.80$.

\subsection{Curie temperature of amorphous Nd$_{x}$Fe$_{1-x}$ alloys}

\begin{figure}[tp]
\begin{center}
\includegraphics[width=6.5 cm]{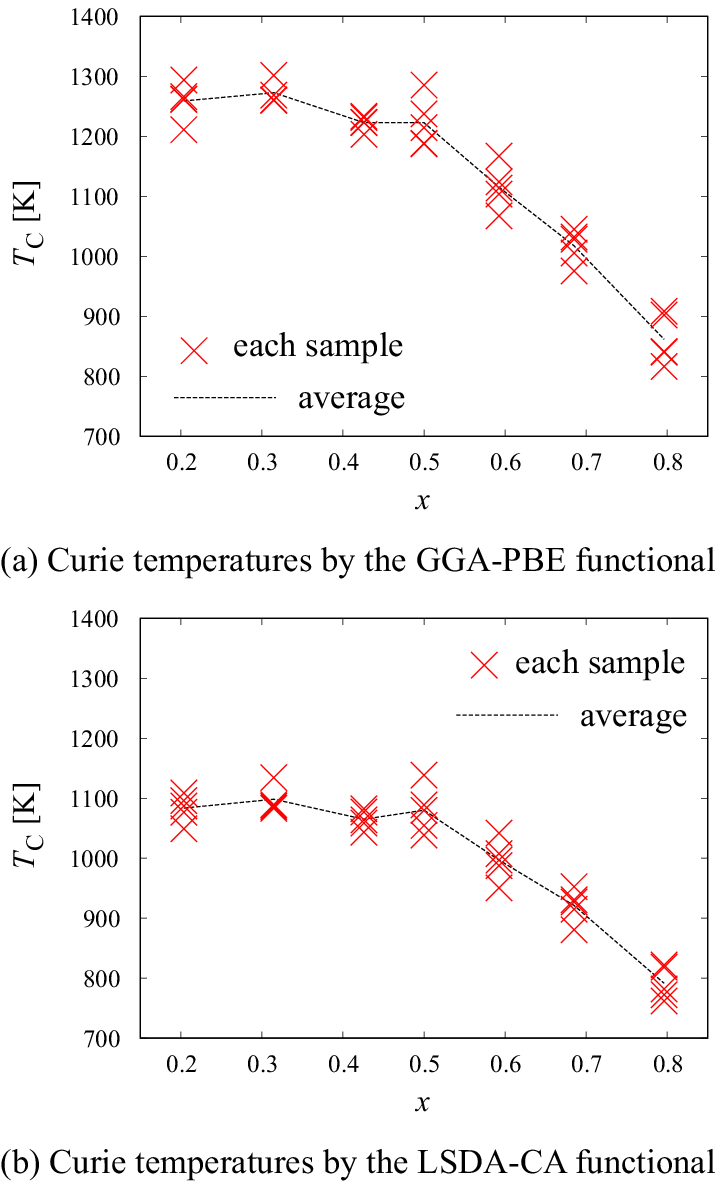}
\end{center}
\caption{
Curie temperatures of amorphous Nd$_{x}$Fe$_{1-x}$ alloys as a function of $x$, calculated within the mean field approximation, by (a) the GGA-PBE and (b) the LSDA-CA functionals.
The red crosses represent the values of different configurations, and the black dashed line represents the average value of different samples with the same composition.
}
\label{fig:curie}
\end{figure}

We evaluated the Curie temperatures of amorphous Nd$_{x}$Fe$_{1-x}$ from the exchange coupling constants using the mean-field approximation.
Within the mean-field approximation, the Curie temperature of a periodic system can be calculated as the maximum eigenvalue of the matrix $\bm{\Theta}$, whose elements can be written as 
\begin{eqnarray}
\Theta_{ij}
&=
 &\frac{2}{3 k_{\mathrm{B}}}
  \sum_{\mathbf{R}\ne \mathbf{0}}J_{\mathbf{0}i,\mathbf{R}j},
\label{eq:curie}
\end{eqnarray}
where $J_{\mathbf{0}i,\mathbf{R}j}$ denotes the exchange coupling constants of atom $i$ at cell $\mathbf{0}$ and atom $j$ at cell $\mathbf{R}$.

Figure \ref{fig:curie}(a) shows the Curie temperatures of the amorphous Nd$_{x}$Fe$_{1-x}$ alloys for different compositions as well as all the examined samples for a given composition.
The maximum Curie temperature was found to be about 1300 K, which is close to the Curie temperature of bcc Fe calculated within the mean-field approximation.
The Curie temperature drops nonlinearly with increasing $x$ and reaches about 850 K at $x=0.80$.
It is also noteworthy that the calculated Curie temperatures depend on the choice of exchange-correlation functional.
Figure \ref{fig:curie}(b) shows the Curie temperatures for the same systems calculated with the exchange correlation functional by D. M. Ceperley and B. J. Alder \cite{LSDA-CA,LSDA-CA_2} with the local spin density approximation (LSDA), namely the LSDA-CA functional.
In this case, the maximum and minimum Curie temperatures are approximately 1150 K and 750 K, respectively.

These results suggest that amorphous Nd--Fe alloys exhibit ferromagnetism even for Nd-rich compositions.
Although this contradicts a naive expectation that the magnetism of an alloy is determined by the ratio of ferromagnetic and paramagnetic materials in the composition, the high $T_{\mathrm{C}}$ of Nd--Fe alloys are also reported in other recent studies.
For example, a recent experimental study reports the ferromagnetism in amorphous Nd$_{x}$Fe$_{1-x}$ alloys for the compositions with $x<0.70$ \cite{A_Sakuma_2015}.
Moreover, the high $T_{\mathrm{C}}$ of fluorite Nd$_{2}$Fe alloys based on the first-principles calculation are also reported in Ref.~\cite{Y_Ainai_2020}, which reported $T_{\mathrm{C}}$ to be 585 K for $x = 0.67$ within the LSDA functional proposed by Moruzzi, Janak, and Williams \cite{V_L_Moruzzi_1978}.
The difference of calculated Curie temperature between Ref.~\cite{Y_Ainai_2020} and our study can be explained partly by the difference of geometry because the distance between the nearest neighbor Fe--Fe pairs is 4.9 \AA\ in the fluorite Nd$_{2}$Fe while the average of the distances between neighboring Fe--Fe pairs is approximately 2.5 \AA\ in our study. 

However, we still have to be careful, as the ferromagnetic GB phase does not account for the coercivity enhancement of permanent magnet.
It is pointed out that the segregation of Cu at the interface of the main and GB phases enhances the magnetic insulation of main phase grains, both experimentally \cite{H_Sepehri_Amin_2012} and theoretically for crystalline Nd--Fe alloy \cite{Y_Ainai_2020}.
The effects of additional Cu and other elements to the amorphous Nd--Fe alloy have to be investigated further in the future.

\subsection{Analysis of averaged features using the Gabriel graph}

\begin{figure}[t]
\begin{center}
\includegraphics[width=7.5 cm]{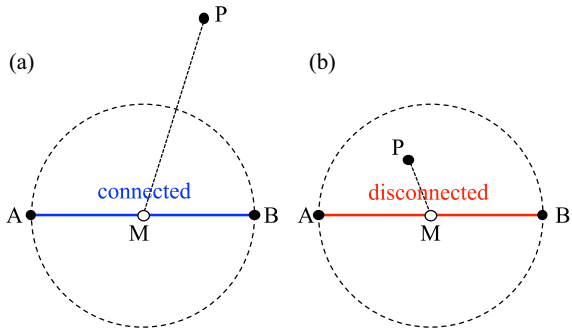}
\end{center}
\caption{
Schematics for the construction scheme of Gabriel graph. (a) 
\rev{
Connect 
}
the points A and B when no other points are found in smaller distances from the center M than A or B. (b) Remove the edge between points A and B when points are found at smaller distances from the center M other than A or B.
}
\label{fig:gabriel}
\end{figure}

To investigate the effect of surrounding atoms on $J_{ij}$, we performed combined analyses of structure and exchange coupling using the Gabriel graph \cite{K_R_Gabriel_1969}.
Gabriel graph provides a scheme to construct a network for a set of points in a metric space, by deciding to connect the two points or not from the center between the two points (see Figure \ref{fig:gabriel}).
We have demonstrated that the Gabriel graph is suitable to describe nearest neighbor networks in amorphous systems in Ref.~\cite{A_Terasawa_2018}.

\begin{figure}[t]
\begin{center}
\includegraphics[width=6.0 cm]{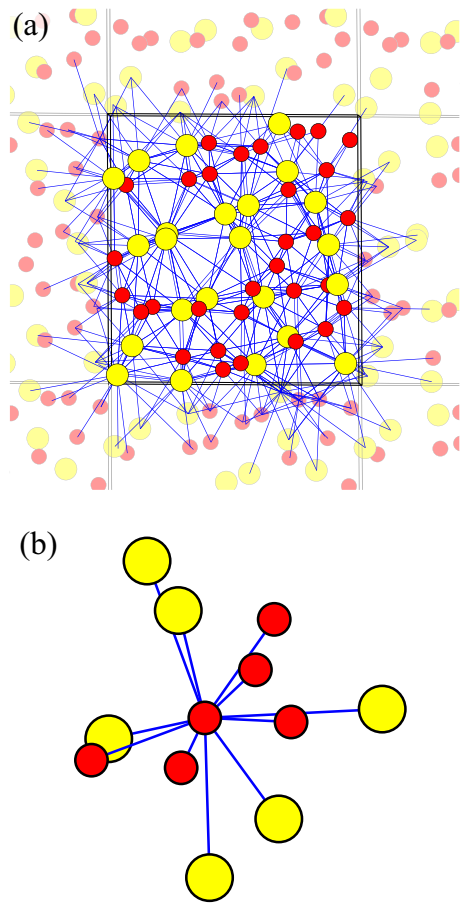}
\end{center}
\caption{
(a) Gabriel graph constructed from one of the examined samples of Nd$_{0.42}$Fe$_{0.58}$ amorphous alloy.
The red and yellow circles correspond to the Fe and Nd atoms, and the blue lines correspond to the edges of the Gabriel graph.
The atoms outside the unit cell are indicated by the pale colors.
(b) The partial Gabriel graph connected to atom 1.
}
\label{fig:ggraph_42}
\end{figure}

Figure \ref{fig:ggraph_42}(a) shows the Gabriel graph constructed from one of the examined samples of Nd$_{0.42}$Fe$_{0.58}$ amorphous alloy.
It is possible to see in the figure that the network of neighboring atoms are constructed by the edges of the Gabriel graph, which are indicated by blue lines.
In the following analysis, we define the pairs of neighboring atoms as the atom pairs having Gabriel graph edges in between.

\begin{figure*}[t]
\begin{center}
\includegraphics[width=17 cm]{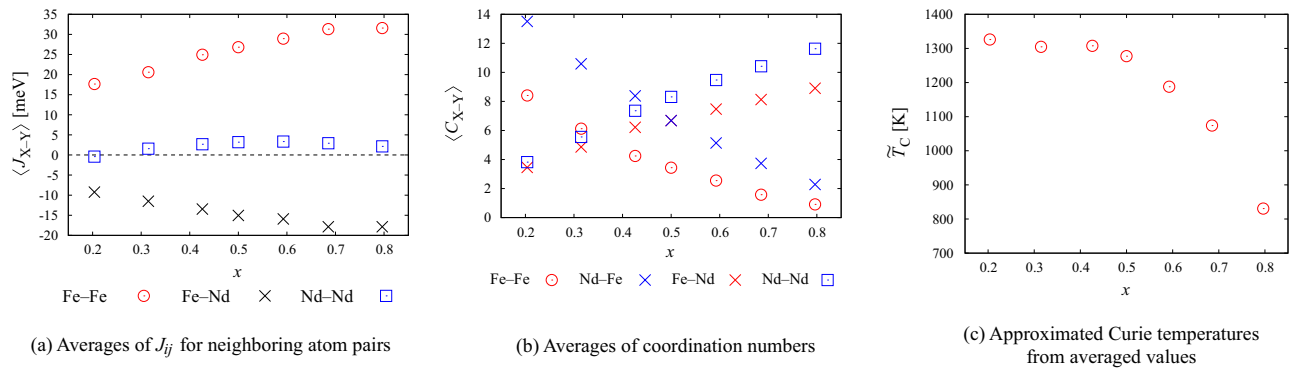}
\end{center}
\caption{
Averaged values of (a) exchange coupling constants and (b) coordination numbers for the edges of the Gabriel graphs. (c) Approximated Curie temperature calculated from the averaged values.
}
\label{fig:graph_avg}
\end{figure*}

Using the Gabriel graph, it is possible to analyze the relationship between the exchange coupling constants and the structural characters of the systems.
As the simplest analysis, we calculated the mean value of $J_{ij}$ for pairs of neighboring atoms.
In Fig.~\ref{fig:graph_avg}(a), we show the averages of $J_{ij}$ for different element pairs.
An important feature that can be seen clearly in the graph is the strong $x$ dependence of the averages of $J_{ij}$ between Fe--Fe and Fe--Nd.

Another important character extracted from the graph analysis is the coordination structure. 
Fig \ref{fig:ggraph_42}(b) shows the partial Gabriel graph connected to atom 1. The number of neighboring Fe atoms to atom 1 is counted as 5, and the number of neighboring Nd atoms to atom 1 is counted as 6. In our graph analysis, we defined those numbers as the coordination numbers.
The averaged coordination numbers are shown in Fig.~\ref{fig:graph_avg}(b) for $J_{ij}$ between different elemental pairs.
Here, ${C}_{\mathrm{X-Y}}$ represents the number of neighboring atoms of species Y around an atom of element X.

Given the averaged exchange $\langle{J}_{\mathrm{X-Y}}\rangle$ and averaged coordination numbers $\langle{C}_{\mathrm{X-Y}}\rangle$, it is possible to evaluate the approximated Curie temperature $\tilde{T}_{\mathrm{C}}$ from the averaged values, as the maximum eigenvalue for the approximated form of Eq.~(\ref{eq:curie}):
\begin{eqnarray}
\bm{\tilde{\Theta}}
&=
 &\frac{2}{3 k_{\mathrm{B}}}
  \!\left(\!\!\begin{array}{cc}
   \langle{C}_{\mathrm{Fe-Fe}}\rangle\langle{J}_{\mathrm{Fe-Fe}}\rangle
   &\langle{C}_{\mathrm{Fe-Nd}}\rangle\langle{J}_{\mathrm{Fe-Nd}}\rangle \\
   \langle{C}_{\mathrm{Nd-Fe}}\rangle\langle{J}_{\mathrm{Fe-Nd}}\rangle
   &\langle{C}_{\mathrm{Nd-Nd}}\rangle\langle{J}_{\mathrm{Nd-Nd}}\rangle \\   
  \end{array}\!\!\right).
\label{eq:curie_2}
\end{eqnarray}
Figure \ref{fig:graph_avg}(c) shows the approximated Curie temperature $\tilde{T}_{\mathrm{C}}$ as a function of $x$, calculated from $\langle{J}_{\mathrm{X-Y}}\rangle$ and $\langle{C}_{\mathrm{X-Y}}\rangle$. Comparing Fig.~\ref{fig:graph_avg}(c) with Fig.~\ref{fig:curie}, we can see that the approximated Curie temperature is similar to that of determined from all the $J_{ij}$ values, except for a small deviation in the positive direction.
This implies that the Curie temperatures of the amorphous Nd$_{x}$Fe$_{1-x}$ alloys are mostly dominated by the exchange coupling of neighboring pairs defined by the Gabriel graph.
We also suggest that the deviation of $\tilde{T}_{\mathrm{C}}$ from $T_{\mathrm{C}}$ may come from the long-range terms of Fe--Fe pairs, which tend to have negative values.

\begin{figure*}[t]
\begin{center}
\includegraphics[width=14 cm]{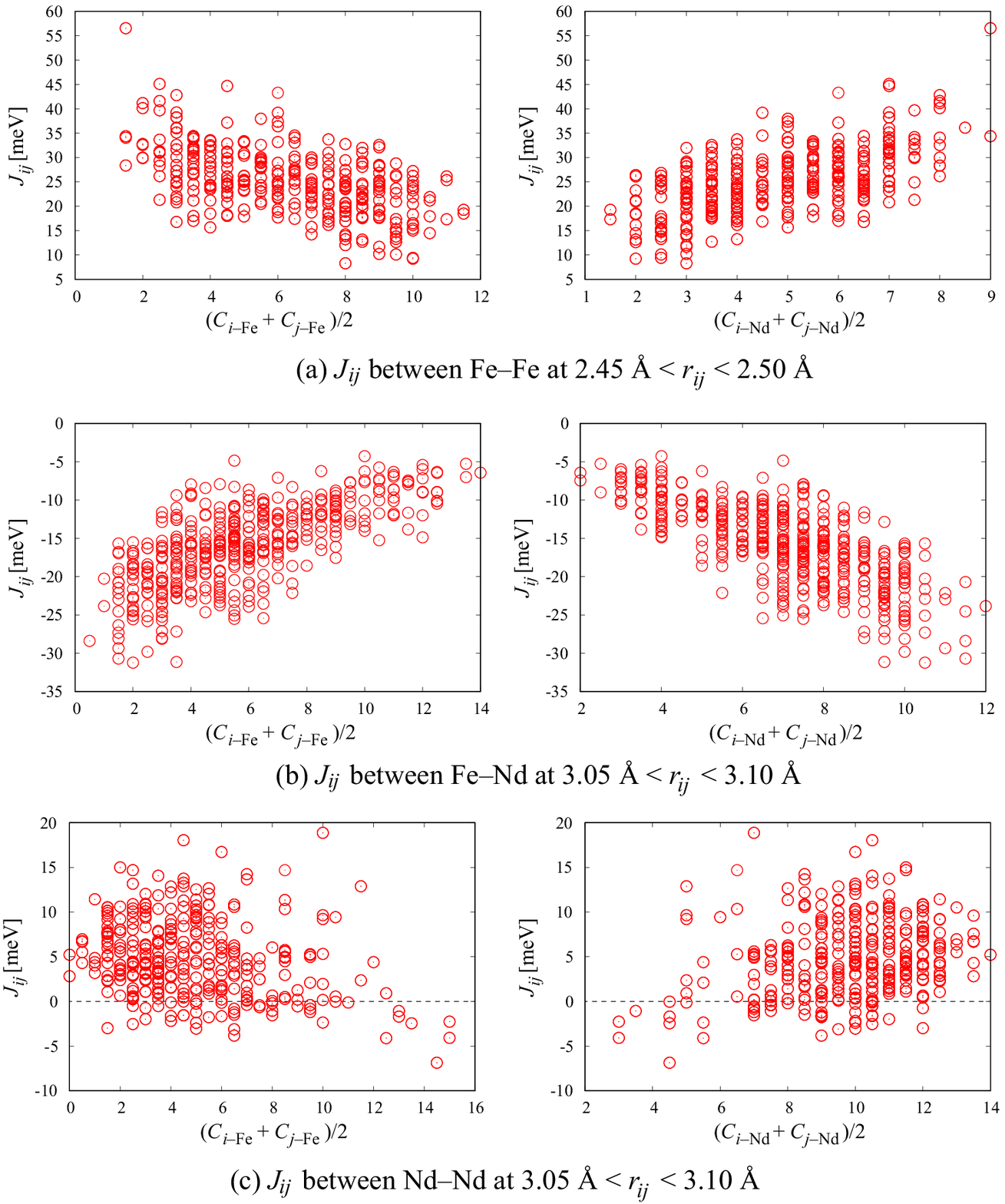}
\end{center}
\caption{The exchange coupling constants $J_{ij}$ as a function of coordination numbers, between (a) Fe--Fe, (b) Fe--Nd, and (c) Nd--Nd of all the examined samples.
In each panel, the sampled range of atomic distances is constrained to the region where the largest number of pairs of neighboring atoms is found.
As the variables in the graphs, we used the mean value $(C_{i-\mathrm{X}}+C_{j-\mathrm{X}})/2$ for atoms $i$ and $j$, where $C_{i-\mathrm{X}}$ corresponds to the number of neighboring atoms of species X to atoms $i$.
}
\label{fig:C_vs_J}
\end{figure*}

\subsection{Local environment analysis using the Gabriel graph}

In addition to the analysis of averages described above, it is possible to analyze the relationship between the exchange coupling constants and the local environment using the coordination numbers for each atom.
Figure \ref{fig:C_vs_J} shows the exchange coupling constants $J_{ij}$ as a function of coordination numbers, between different elemental pairs and a small range of atomic distances.
Here, $C_{i-\mathrm{X}}$ denotes the number of neighboring atoms of element X to atom $i$. We used the mean values for atoms $i$ and $j$ as the variables in Fig.~\ref{fig:C_vs_J}.
In all the three, the top, middle and bottom panels of Fig.~\ref{fig:C_vs_J}, we show the $J_{ij}$ as a function of the number of surrounding Fe atoms in the left graph and $J_{ij}$ as a function of the number of surrounding Nd atoms in the right graph.
In Figs.~\ref{fig:C_vs_J}(a) and (b), it is possible to see a clear tendency that $J_{ij}$ becomes weaker with the increasing number of surrounding Fe atoms, and becomes stronger with the increasing number of Nd atoms. In comparison, the tendency is rather ambiguous in Fig.~\ref{fig:C_vs_J}(c).
This is another evidence that the exchange coupling constants between two atoms are suppressed by surrounding Fe atoms and/or enhanced by surrounding Nd atoms.

\section{Summary}

In this paper, we examined the exchange coupling constants $J_{ij}$ of amorphous Nd$_{x}$Fe$_{1-x}$ alloys in the framework of spin-dependent density functional calculation.
We observed a strong fluctuation in the exchange coupling constants when plotted against the atomic distances.
By comparing the $J_{ij}$ curves among different compositions, we found that the shapes of $J_{ij}$ curves are modified by changing the composition.
This resulted in the nonlinear dependence of Curie temperature as a function of composition ratio, and the Curie temperature remained large even for Nd-rich 
compositions.
To evaluate the effect of the surrounding environment on $J_{ij}$, we performed combined analyses of structure and exchange coupling using the Gabriel graph.
It was found that the averages of $J_{ij}$ between Fe--Fe and Fe--Nd increase with increasing the number of surrounding Nd atoms.
We also investigated $J_{ij}$ as a function of coordination numbers, and found that the exchange couplings become stronger with increasing Nd concentration.
The enhancement of $J_{ij}$ under the circumstance of large Nd density seen in our study may account for the ferromagnetism of Nd--Fe amorphous alloy observed in the experiments.
The graph analysis of the exchange coupling proposed in this study can open up the new schemes to understand complicated magnetic interactions in amorphous systems.

\begin{acknowledgments}
The authors thank Taisuke Ozaki, Hisazumi Akai, Munehisa Matsumoto, Yuta Toga, Sonju Kou, and Yuta Ainai for fruitful discussions and valuable comments.

This work was supported in part by MEXT, Japan as Program for Promoting Researches on the Supercomputer Fugaku, DPMSD, the Elements Strategy Initiative Project (ESICMM, Grant No.~JPMXP0112101004) under the auspices of MEXT, as well as JSPS-KAKENHI Grant No.~17K04978. 
Some of the calculations were performed using the supercomputers at ISSP, The University of Tokyo, and TSUBAME, Tokyo Institute of Technology, as well as the K computer, RIKEN Project Nos.~hp180206, and hp190169).
\end{acknowledgments}

\end{document}